\documentclass[conference, 10pt]{IEEEtran}
\IEEEoverridecommandlockouts
\usepackage{cite}
\usepackage{amsmath}
\usepackage{amsfonts}
\usepackage{bm}  % bold Greek symbols
\usepackage{graphicx}
\usepackage{hyperref}
\usepackage{amsthm} % Denifition, Theorem, Proposition, etc
\usepackage{authblk} % author group's marks

\newtheorem{definition}{Definition}
\newtheorem{proposition}{Proposition}

\def\BibTeX{{\rm B\kern-.05em{\sc i\kern-.025em b}\kern-.08em
    T\kern-.1667em\lower.7ex\hbox{E}\kern-.125emX}}
\makeatletter
\def\endthebibliography{%
	\def\@noitemerr{\@latex@warning{Empty `thebibliography' environment}}%
	\endlist
}

\begin{document}
\title{\huge A Lightweight Human Pose Estimation Approach for Edge Computing-Enabled Metaverse with Compressive Sensing}
\author{Nguyen Quang Hieu}
\author{Dinh Thai Hoang}
\author{Diep N. Nguyen}
\affil{School of Electrical and Data Engineering, University of Technology Sydney, Australia}
\maketitle
\begin{abstract}
The ability to estimate 3D movements of users over edge computing-enabled networks, such as 5G/6G networks, is a key enabler for the new era of extended reality (XR) and Metaverse applications. Recent advancements in deep learning have shown advantages over optimization techniques for estimating 3D human poses given spare measurements from sensor signals, i.e., inertial measurement unit (IMU) sensors attached to the XR devices. However, the existing works lack applicability to wireless systems, where transmitting the IMU signals over noisy wireless networks poses significant challenges. Furthermore, the potential redundancy of the IMU signals has not been considered, resulting in highly redundant transmissions. In this work, we propose a novel approach for redundancy removal and lightweight transmission of IMU signals over noisy wireless environments. Our approach utilizes a random Gaussian matrix to transform the original signal into a lower-dimensional space. By leveraging the compressive sensing theory, we have proved that the designed Gaussian matrix can project the signal into a lower-dimensional space and preserve the Set-Restricted Eigenvalue condition, subject to a power transmission constraint. Furthermore, we develop a deep generative model at the receiver to recover the original IMU signals from noisy compressed data, thus enabling the creation of 3D human body movements at the receiver for XR and Metaverse applications. Simulation results on a real-world IMU dataset show that our framework can achieve highly accurate 3D human poses of the user using only $82\%$ of the measurements from the original signals. This is comparable to an optimization-based approach, i.e., Lasso, but is an order of magnitude faster.
\end{abstract}


\section{Introduction}
The past few years witnessed an ever-increasing demand for virtual reality (VR) and augmented reality (AR) applications over mobile networks (5G networks and upcoming 6G networks). Following active academic research of VR/AR technologies \cite{siriwardhana2021survey}, industrial leading organizations, such as 3GPP groups, have shown early steps in incorporating VR/AR into the next network generations, thus enhancing extended reality (XR) capabilities for Metaverse applications \cite{5gamericas2024}. The 3D nature of these XR and Metaverse applications brings new challenges to network design in which conventional approaches are not sufficient. Such new challenges are field-of-view transmissions for 360-degree videos/images \cite{guo2023federated}, the capacity overhead of uplink transmission \cite{jiang2023qoe}, and motion sickness of changing scene resolution \cite{fernandes2016combating}, to name a few. In these challenges, one of the most recognized patterns is changing user body movements, e.g., changing head orientation, acceleration, and height, which results in sudden updates in the virtual 3D environment. The service provider, e.g., an edge computing server, has to quickly adapt to, or predict the changes of the user, and then update the user's QoE accordingly, e.g., updating the field-of-view's position \cite{guo2023federated}, resolution \cite{fernandes2016combating}, or allocating more physical resources to that frames within the new field-of-view \cite{5gamericas2024}. 

The approaches in \cite{5gamericas2024, fernandes2016combating, jiang2023qoe, guo2023federated}, and in the references therein, usually focus on the head orientation of the user, which is similar to conventional mobile phone user when changing the phone's direction/orientation can degrade the received signal strength. However, in emerging Metaverse services over wireless networks, such as real-time gaming and virtual interactive events, having only head orientation is not enough. In such services, having the full 3D movements of users is usually much more valuable \cite{siriwardhana2021survey}. However, estimating the 3D body movements of users is a much more challenging task, given noisy wireless environments and existing hardware VR/AR devices' constraints. The literature works on estimating 3D human movements, or 3D human poses, suggest that leveraging calibrated camera systems gives the best performance, i.e., estimation accuracy. However, these camera systems are limited to indoor scenarios and are prone to lightning conditions. Furthermore, capturing user activities via camera usually reveals vulnerable privacy issues, especially when these captured images/videos are transmitted via wireless networks. Thus, camera-based approaches are not sufficient for scalable solutions of 3D human pose estimation over wireless Metaverse systems. 

Fortunately, recent approaches leverage another source of information from the built-in hardware devices within almost every VR/AR user device, which is the inertial measurement unit (IMU) sensor. The IMU sensors, which include accelerometers, gyroscopes, and magnetometers, are integrated into devices such as VR/AR headsets and joystick controllers to measure their orientation, acceleration, altitude, and angular velocity. More complex IMU systems may have a higher number of IMU sensors attached, for example, on a body suite \cite{xsens2023}. The literature works in \cite{von2017sparse, huang2018deep, yi2022physical, xsens2023} suggest that the entire 3D body movement can be accurately constructed with sparse measurements from IMU sensors, ranging from 3 to 17 sensors. These approaches can be divided into two categories that are (i) optimization-based approach \cite{von2017sparse, xsens2023} and (ii) learning-based approaches \cite{huang2018deep, yi2022physical}. Optimization approaches apply Kalman Filter \cite{xsens2023} or exponential mapping \cite{von2017sparse} to approximate the 3D poses of a given human kinematic body model \cite{loper2023smpl}.  Unlike optimization approaches, learning-based approaches can achieve real-time estimation based on pre-trained deep learning models. In return, the learning-based approaches may suffer higher uncertain performance and require large datasets due to the training nature of the deep learning models. Additionally, a physic-aware optimizer can also be used to enhance the tracking accuracy with advantages in using a longer prediction time window \cite{yi2022physical}.

Despite recent advancements in enhancing the precision of 3D human pose reconstruction within VR/AR environments, the implementation of these technologies across wireless networks, such as 5G and 6G systems, remains inadequate. Deploying these approaches over wireless networks like 5G/6G systems is challenging because wireless channels are prone to noise or interference, which can significantly impact the transmitted data. Existing works in \cite{von2017sparse, huang2018deep, yi2022physical, xsens2023} overlook the presence of noise or damaged packets in the received IMU data. This can lead to degraded reconstruction accuracy and poor user experience in the Metaverse environment. Additionally, these approaches often use computationally expensive methods such as deep learning without considering the potential redundancy of the IMU data before transmitting it to the receiver. While data compression approaches can help to remove the potential redundancy in the IMU data, these compression approaches, e.g., entropy coding, are computationally expensive for the IMU data acquisition and transmission. 

To address these limitations, we propose a novel approach that utilizes a random Gaussian matrix following the compressive sensing theory,  i.e., the Gaussian matrix that satisfies the set-restricted Eigenvalue condition \cite{bora2017compressed}, to enable lightweight transmission of the compressed IMU data at the transmitter. At the receiver, we develop a deep generative model to reconstruct the original IMU signal from the noisy and compressed IMU data transmitted over a wireless channel. The novelty of our proposed framework is twofold. First, we design a new measurement matrix for down-sampling the original IMU signal into a lower dimensional space via linear projection, subject to a transmission power constraint. Second, we leverage a deep generative model at the receiver to recover the original IMU. This system model benefits the transmitter via a lightweight linear transform of the IMU signal, while the computationally expensive task, i.e., reconstructing the original IMU signal, is performed at the receiver, e.g., an edge computing server, with more computing power. Our system model is suitable for edge computing-enabled Metaverse applications with lightweight IMU transmitters. We will show later that our framework can achieve competitive signal reconstruction accuracy by using fewer measurements from the original signals (i.e., $82\%$ of original signals), but with significantly lower reconstruction latency, compared to the optimal solution obtained by an optimization-based approach.

\section{System Model and Preliminaries}
\begin{figure}
\centering
\includegraphics[width=0.6\linewidth]{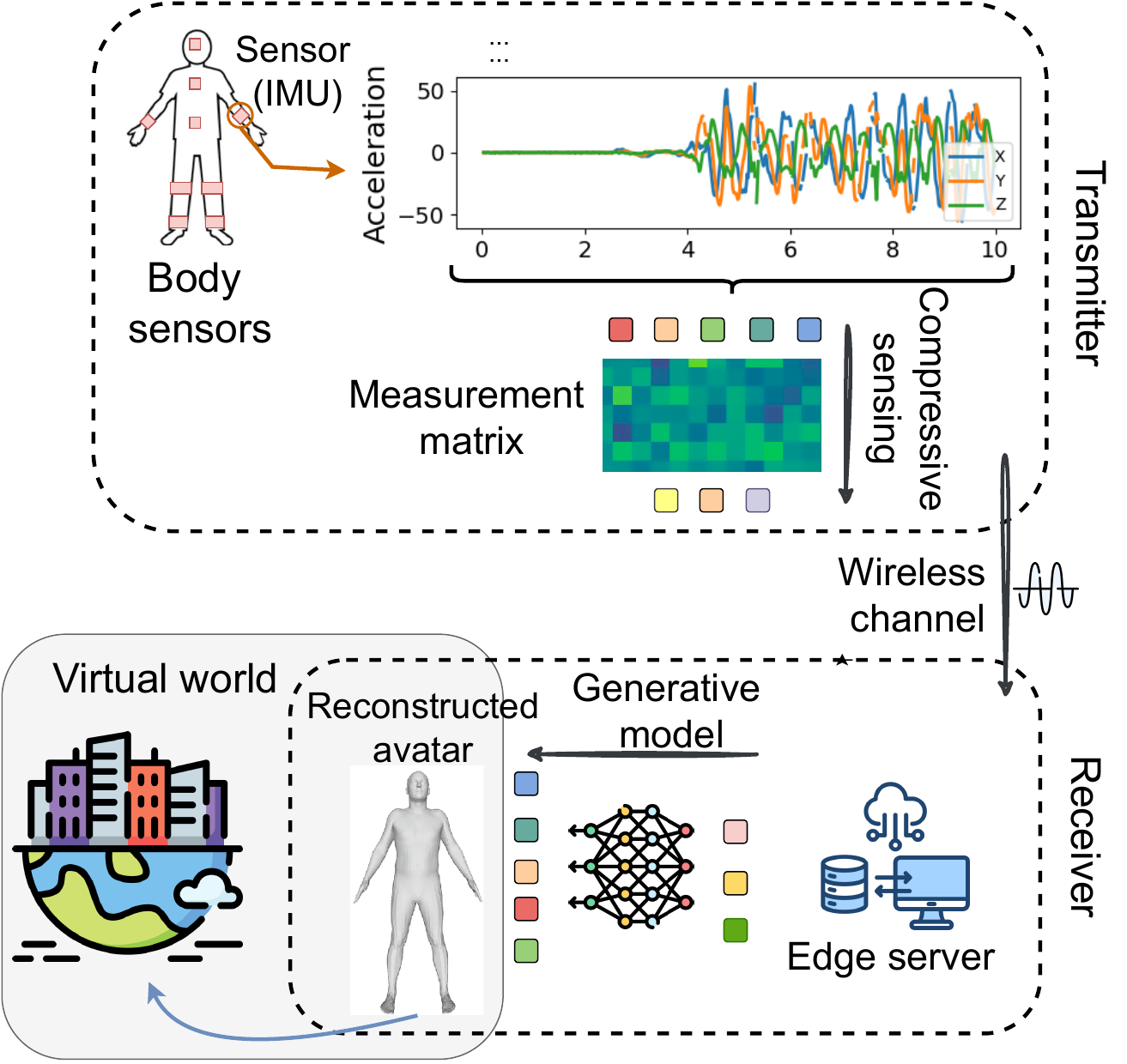}
\caption{The proposed system model. A linear projection is performed at the transmitter with compressive sensing. The down-sampled IMU signals are transmitted via a wireless channel. The receiver empowered with a generative model recovers the original IMU signals from the noisy compressed data. Finally, the reconstructed avatar from the recovered IMU signals can be used in the Metaverse's virtual world.}
\label{fig:system-model}
\end{figure}

The considered system model is illustrated in Fig.~\ref{fig:system-model}. At the transmitter side, the user is equipped with a set of IMU sensors, e.g., Xsens' IMU sensors \cite{xsens2023}. The sensors are usually placed on the body joints, e.g., head, shoulders, wrists, elbows, waist, knees, and ankles \cite{huang2018deep, xsens2023}. Each IMU sensor produces a sequence of signals measuring the information of the body joint (e.g., see Fig.~\ref{fig:fft-imu}). By combining the information from the sensors, we can fully recoverer the 3D body poses of the user up to a certain accuracy level with respect to the number of sensors being used \cite{von2017sparse}. 
\subsection{Compressive Sensing}
After having a fixed sequence of $n$ ($n > 0$) data points about the acceleration and orientation of all IMU sensors, we have a finite original signal $\mathbf{x}^* \in \mathbb{R}^n$ (illustrated as five multi-color squares within the transmitter in Fig.~\ref{fig:system-model}). The original vector is then down-sampled by a linear projection with the measurement matrix $\mathbf{A} \in \mathbb{R}^{m \times n}$, where $m$ is the number of measurements, or the length of the down-sampled vector, $\mathbf{y} = \mathbf{Ax}^*$. After that, the vector $\mathbf{y}$ will be transmitted to the receiver over a wireless channel. 
%In this work, we do not consider the quantization and modulation of the signal before transmission as we will later show that the receiver empowered with a deep generative model has the ability to recover the original signal with high accuracy. The error bound and the uniqueness of the recovered signal will be further discussed in later sections. 
Assuming a Gaussian channel with additive white noise, the received signal is $\mathbf{\hat{y}} = \mathbf{Ax}^* + \bm{\eta}$, where $\bm{\eta} \in \mathbb{R}^m$ is the Gaussian noise vector with zero mean and $\sigma_N$ standard deviation, i.e., element $\eta_i$ ($i=1, 2, \ldots, m$) of $\bm{\eta}$ follows a Normal distribution $\eta_i \sim \mathcal{N}(0, \sigma_N^2)$. Despite the simplicity of the channel model, the white Gaussian channel has general properties that can be extended to other complicated channels, e.g., fading channels. 
In order to reconstruct the original signal $\mathbf{x}^*$ from the noisy measurement $\mathbf{\hat{y}}$, the receiver needs to solve the following optimization problem:
\begin{equation}
\begin{split}
\mathcal{P}_0: & \quad \min_{\mathbf{x}} \|\mathbf{x}\|_1, \\
\text{subject to} & \quad \|\mathbf{Ax} - \mathbf{\hat{y}}\|_2 \leq \|\bm{\eta}\|_2,
\end{split}
\label{eq:l1-min}
\end{equation} 
where the notation $\|\mathbf{x}\|_p$ denotes the $l_p$ norm ($p=0, 1, 2, \ldots$) of the vector $\mathbf{x}$, i.e.,
\begin{equation}
\label{eq:lp-norm}
\|\mathbf{x}\|_p = \Big(\sum_{j=1}^{n} |x_j|^p\Big)^{1/p}.
\end{equation}

The objective of $\mathcal{P}_0$ is to find the valid vector $\mathbf{x}$ that minimizes the sum of the absolute value of the elements $x_j$ ($j = 1, 2, \ldots, n$), i.e., $\|\mathbf{x}\|_1 = \sum_{j=1}^{n}|x_j|$, subject to a reconstruction error constraint. Intuitively, minimizing the sum $\|\mathbf{x}\|_1$ encourages the vector $\mathbf{x}$ to be ``sparse" to satisfy the reconstruction error constraint. In particular, the problem $\mathcal{P}_0$ forms an underdetermined system, meaning that there will be multiple solutions to the system if no additional assumptions have been made. To ensure the unique recovery of the signal, compressive sensing theory relies on two main assumptions as follows. First, the signal $\mathbf{x}^*$ must have the sparsity property so that $\mathbf{x}^*$ will be ``compressible". Second, the matrix $\mathbf{A}$ must satisfy specific conditions, such as the Restricted Eigenvalue Condition (REC), which guarantees that the linear projection of $\mathbf{x}^*$ into a lower dimensional space is recoverable \cite{foucart2013invitation}. 

\begin{figure}
\centering
\includegraphics[width=0.6\linewidth]{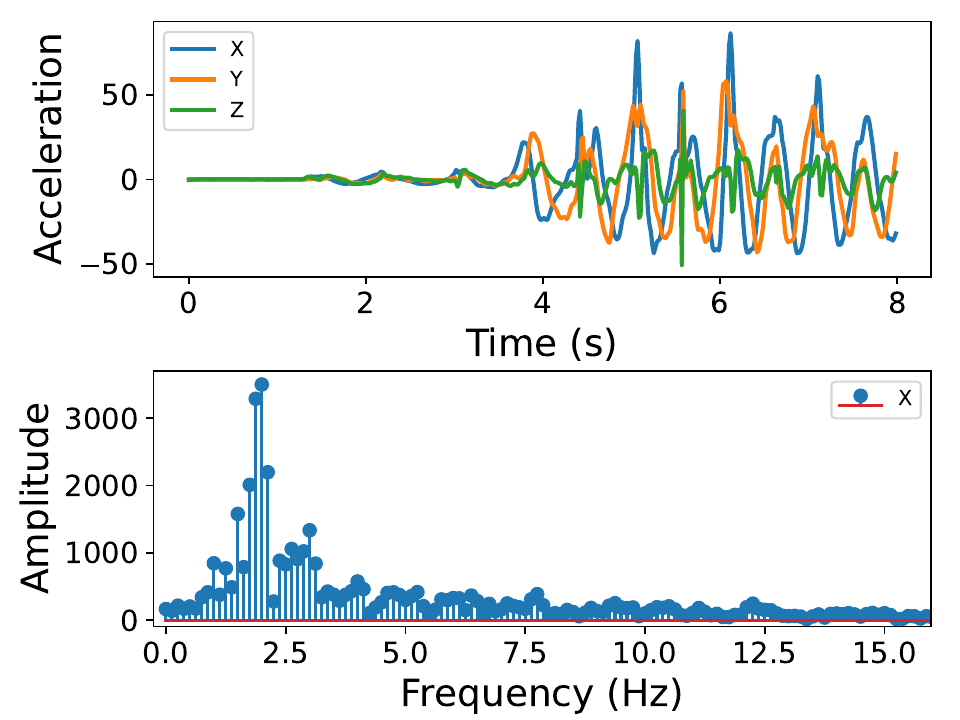}
\caption{Top figure: Acceleration reading from an IMU sensor placed on the left wrist of the user in dataset \cite{huang2018deep}. Bottom figure: Fast Fourier Transform (FFT) of the acceleration data.}
\label{fig:fft-imu}
\end{figure}

\textbf{Sparsity of Vector $\mathbf{x}^*$:} In compressive sensing, the main idea behind compressing the original vector into a lower dimensional space is exploiting the sparsity of the signal. Interestingly, many types of signals in practice are spare or nearly sparse, either in the original form or in transformed domains, e.g., frequency domain. In Fig.~\ref{fig:fft-imu}, we show that the IMU signal from the sensor is $k$-sparse in the frequency domain. We apply the Fast Fourier Transform (FFT) for the acceleration reading from one IMU sensor in dataset \cite{huang2018deep} along its x-axis. The FFT reveals that a few low-frequency coefficients have dominant values. Thus, the potential redundancy of the IMU signal can be approximated by preserving the $k$ largest coefficients and assuming the other coefficients are zero. Note that we use the FFT in Fig.~\ref{fig:fft-imu} for illustrative purposes only, as the FFT may require high computation cost at the transmitter. Next, we discuss how compressive sensing exploits the sparsity of the signal, compresses it, and recovers the original signal at the receiver via optimization or learning approaches.

\textbf{Properties of Matrix $\mathbf{A}$}: The second component of compressive sensing is the use of random matrices for transforming the original vector into a lower dimensional space. It is known that random matrices like Gaussian matrices, i.e., entries of the matrix are drawn from a Normal distribution \cite{foucart2013invitation}, can ensure the recoverability of the compressed signal. The main idea behind using random matrices is that these matrices preserve sparse signals, meaning that the coefficients of the transformed signal are far from the null space, making it possible to recover such signals with high accuracy. The definition of this property is the Restricted Eigenvalue Condition (REC), which is defined as follows.

\begin{definition}[Restricted Eigenvalue Condition (REC)]
\label{eq:rec}
Let $S_k \subset \mathbb{R}^n$ be the set of $k$-sparse vectors. For some parameter $\gamma > 0$, a matrix $\mathbf{A} \in \mathbb{R}^{m \times n}$ is said to satisfy \text{REC}($k, \gamma$) if $\forall \mathbf{x} \in S_k$,
\begin{equation*}
\|\mathbf{Ax}\|_2 \geq \gamma \|\mathbf{x}\|_2. 
\end{equation*}
\end{definition}

The REC property ensures that the smallest singular values of certain submatrices of the sensing matrix are bounded away from zero. This condition is crucial for guaranteeing the stability and accuracy of sparse signal recovery algorithms \cite{raskutti2010restricted}. The algorithms to recover the signals with matrix $\mathbf{A}$ satisfying REC property are elaborated as follows.

\textbf{Signal Recovery:}
As observed from equations (\ref{eq:l1-min}) and (\ref{eq:lp-norm}), the $l_1$ norm of $\mathbf{x}$ is a convex function, thus $\mathcal{P}_0$ is a convex optimization problem. To avoid the computational cost of convex optimization solvers, especially when $n$ is sufficiently large, the problem $\mathcal{P}_0$ can be transformed into solving $\mathcal{P}_1$, for some parameter $\tau \geq 0$ \cite{foucart2013invitation}:
\begin{equation}
\label{eq:transformed-l2-min}
\begin{split}
\mathcal{P}_1: & \quad \min_{\mathbf{x}} \|\mathbf{Ax - \hat{y}}\|_2, \\
\text{subject to} & \quad \|\mathbf{x}\|_1 \leq \tau.
\end{split}
\end{equation}
With the transformed problem, $\mathcal{P}_1$ can be effectively solved with an optimization solver, such as Lasso (Lease absolute shrinkage and selection operator) \cite{foucart2013invitation, bora2017compressed}. The output of Lasso is equivalent to solving the Lagrangian of the problem $\mathcal{P}_1$, for some parameter $\lambda \geq 0$ \cite[Equation 3.1]{foucart2013invitation}:
\begin{equation}
\min_{\mathbf{x}} \|\mathbf{Ax - \hat{y}}\|_2^2 + \lambda \|\mathbf{x}\|_1.
\label{eq:lasso-lagrangian}
\end{equation}

It can be observed in (\ref{eq:lasso-lagrangian}) that the penalty $\lambda \|\mathbf{x}\|_1$ enforces sparsity in the $l_1$ norm space of the vector $\mathbf{x}$. Thus, the $k$-sparse property of the original vector has an impact on the performance of the Lasso solver. In addition, the REC property of $\mathbf{A}$ guarantees that the compressed signal $\mathbf{\hat{y}}$ is recoverable. In this work, we will not explicitly analyze the sparsity of the IMU signal (e.g., through Fourier and/or Wavelet transforms), as this can cause extra computational cost and poor recovery performance in certain settings, e.g., when the linear measurement is not sufficiently large \cite{dhar2018modeling}. Instead, we explore the use of generative models, such as variational auto-encoders (VAEs) \cite{kingma2013auto} and generative adversarial networks (GANs), to learn the underlying distribution of complex and high-dimensional data, thus eliminating the need for conventional sparsity assumptions in conventional compressive sensing methods.

\subsection{Compressive sensing with Generative Models}
Generative modeling is a class of machine learning that involves the process of modeling the underlying distribution given some input dataset.  In particular, given some datasets $D = \{\mathbf{x}_i\}_{i=1}^{N}$, where each data point $\mathbf{x}_{i} \in \mathbb{R}^n$ is usually a $n$-dimensional vector. Generative modeling learns the generative probability (or the forward probability) distribution $P(D|\bm{\theta})$ that approximates the true data distribution $P(D)$, given some model's parameters $\bm{\theta}$ (e.g., a vector, or a matrix, or a deep neural network). In this work, we are interested in a class of generative model that is the variational-autoencoder (VAE) \cite{kingma2013auto} and use it as our model for approximating the underlying distribution of the IMU signals taken from a real-world dataset \cite{huang2018deep}. Note that the selection of a generative model is flexible to our later algorithm, for example, GANs or generative diffusion models can also be used. The main reason for us to use the VAE is its simplicity in training, evaluating the variational lower bound (which will be discussed later), and generating more dispersed samples over the training data.
To train the VAE, we optimize a variational lower bound on the log-likelihood, i.e., 
\begin{equation}
\label{eq:variational-lower-bound}
\mathcal{L}({\mathbf{x}| \bm{\theta; \phi}}) = \sum_{\mathbf{z}} Q(\mathbf{z}|\mathbf{x}; \bm{\phi}) \log \frac{P(\mathbf{x}| \mathbf{z}; \bm{\theta}) P(\mathbf{z})}{Q(\mathbf{z}|\mathbf{x}; \bm{\phi})},
\end{equation}
where $\bm{\theta}$ is the parameters of the VAE's decoder (i.e., the generative model), $\bm{\phi}$ is the parameters of the VAE's encoder (i.e., the inference model), $\mathbf{z}$ is the latent vector,  $Q(\mathbf{z}|\mathbf{x}; \bm{\phi})$ is the variational posterior, and $P(\mathbf{z})$ is the prior. The added parameters $\mathbf{z}$ and $Q(\mathbf{z}|\mathbf{x}; \bm{\phi})$ are for the variational modeling of the input data, where the prior $P(\mathbf{z})$ is usually assumed to follow a simple distribution, e.g., a multivariate Gaussian distribution. 

As sampling from the posterior is straightforward in VAE due to the choice of latent prior (e.g., Gaussian), the VAE model can be trained with unbiased Monte Carlo estimate of $\mathcal{L}$ in (\ref{eq:variational-lower-bound}) using a gradient decent optimizer, e.g., SGD or Adam, resulting in the unbiased estimation of $\mathcal{L}$ \cite{kingma2013auto}:
\begin{equation}
\label{eq:monte-carlo-elbo}
\mathcal{\hat{L}}(\mathbf{x}|\bm{\theta}; \bm{\phi}) = \log \frac{P(\mathbf{x}|\mathbf{z}; \bm{\theta}) P(\mathbf{z})}{Q (\mathbf{z}|\mathbf{x}; \bm{\phi})},
\end{equation}
where $P(\mathbf{x}|\mathbf{z}; \bm{\theta})$ can be calculated based on the output of the VAE's decoder via an activation function, e.g., a Sigmoid function, and $Q(\mathbf{z}|\mathbf{x}; \bm{\phi})$ can be effectively approximated via a reparameterization trick \cite{kingma2013auto}.
Once the variational lower bound is optimized, we can approximate the true probability density function $P(D)$ of the dataset, i.e., $P(D|\bm{\theta}) \approx P(D)$. 

In the context of compressing sensing, we do not have the full observation of the data points $\mathbf{x}_i$ \cite{foucart2013invitation}. In particular, in our considered setting, the receiver has access to the compressed and noisy observation $\mathbf{\hat{y}} = \mathbf{Ax}^* + \bm{\eta}$, where the original vector is $\mathbf{x}^* = \mathbf{x}_i \in D$. Replacing $\mathbf{x}$ with $\mathbf{y} = \mathbf{Ax}$, the estimation of the variational lower bound in (\ref{eq:monte-carlo-elbo}) can be rewritten as:
\begin{equation}
\mathcal{\hat{L}}(\mathbf{y}|\bm{\theta};\bm{\phi}) = \log \frac{P(\mathbf{Ax}|\mathbf{z};\bm{\theta}) P(\mathbf{z})}{Q(\mathbf{z}|\mathbf{Ax};\bm{\phi})}.
\end{equation}

As we now approximate the probability distribution of the signals via the generative model, we need to define the range of the output generator function, i.e., $S_G = \{G(\mathbf{z}): \mathbf{z} \in \mathbb{R}^k\}$.
The reconstructed signals' range is now transformed into the latent space $\mathbf{z} \in \mathbb{R}^k$ ($k < n$). This means that the measurement matrix $\mathbf{A}$'s properties are no longer sufficient to guarantee the accuracy of the reconstructed signals. Thus, with generative model-based compressive sensing, we now require $\mathbf{A}$ to satisfy the Set-Restricted Eigenvalue Condition (S-REC), which is a more comprehensive version of REC \cite{bora2017compressed}. The S-REC property is defined as follows.
\begin{definition}[Set-Restricted Eigenvalue Condition]
\label{def:s-rec}
Let $S \subset \mathbb{R}^n$, for some parameters $\gamma > 0$ and $\kappa \geq 0$, a matrix $\mathbf{A} \in \mathbb{R}^{m \times n}$ is said to satisfy the $\text{S-REC}(S, \gamma, \kappa)$ if $\forall \mathbf{x}_1, \mathbf{x}_2 \in S$,
\begin{equation*}
\|\mathbf{A}(\mathbf{x}_1 - \mathbf{x}_2)\|_2 \geq \gamma \|\mathbf{x}_1 - \mathbf{x}_2\|_2 - \kappa.
\end{equation*} 
\end{definition}

The S-REC property enables training the VAE as the optimization solver for the compressive sensing problems while ensuring the recoverable property of the compressed signals.

\section{Problem Formulation and Proposed Solution}
With the introduced generative model-based compressive sensing method, which was first explored in \cite{bora2017compressed}, we aim to develop a practical framework for the reconstruction of 3D human pose over wireless channels. Recall that the presence of the wireless channel poses significant challenges to the conventional 3D human pose reconstruction methods, and the conventional generative model-based compressive sensing approaches, e.g., \cite{bora2017compressed} and \cite{dhar2018modeling}, cannot be straightforwardly applied. In particular, the measurement matrices used in \cite{bora2017compressed, dhar2018modeling} cannot guarantee the system constraints, i.e., transmission power constraint per channel use.
Designing the measurement matrix $\mathbf{A}$ at the transmitter to ensure both S-REC property and power transmission is a challenging task that has not been considered in the literature. As we will discuss later, existing works that overlook the presence of additive channel noise or utilize simple and intuitive power normalization schemes, such as $l_2$ normalization, may fail in reconstructing 3D human poses at the receiver. Before going into further analysis, let us define the considered problem with the aforementioned constraints:
\begin{equation}
\label{eq:problem-main}
\begin{split}
\mathcal{P}_2: \quad & \min_{\mathbf{z}} \|\mathbf{A} G(\mathbf{z}) - \mathbf{\hat{y}}\|_2, \\
\text{subject to} \quad & \frac{1}{m} \|\mathbf{y}\|_2 \leq P_T, \\
				 \quad & \|G(\mathbf{z})\|_1 \leq \nu,
\end{split}
\end{equation}
where $G(\mathbf{z})$ is the generator function, i.e., the output of the VAE's decoder given the latent vector $\mathbf{z}$, $P_T$ is the power constraint of the transmitter per channel use, and $\nu \geq 0$ is the $l_1$ constraint of the generator function (similar to $\tau$ in the original dimension of $\mathbf{x}$ in equations (\ref{eq:l1-min}) and (\ref{eq:lasso-lagrangian})).

As seen in problem $\mathcal{P}_2$ from (\ref{eq:problem-main}), the power constraint (\ref{eq:problem-main}b) sets an upper limit on the transmitted signal $\mathbf{y} = \mathbf{Ax}$, while also requiring the lower bound of the S-REC property in Definition \ref{def:s-rec}. To the best of our knowledge, this is the first work that investigates how to design a measurement matrix that satisfies this duo constraint. To address this problem, we first propose a novel measurement matrix $\mathbf{A}$ in Proposition \ref{prop:main} which ensures that $\mathbf{y = Ax}$ satisfies the power constraint (\ref{eq:problem-main}b). Once the power constraint is established, we use the Lagrangian of $\mathcal{P}_2$ as a loss function to train the VAE model similar to the approach in \cite{bora2017compressed, dhar2018modeling}.
The proposed measurement matrix for problem $\mathcal{P}_2$ is as follows.

\begin{proposition}[S-REC with power constraint]
The recovered signal obtained by the generative model-based compressive sensing method under the power constraint $P_T$ is guaranteed to be a unique solution if (i) $\mathbf{A}$ satisfies S-REC property, and (ii) each element $A_{ij}$ (element $j$-th of the $i$-th row) of $\mathbf{A}$ is drawn i.i.d from a Gaussian distribution with zero mean and variance $\sigma_a^2 = \frac{P_T}{n^2 d^2 (d \sigma_x + \mu_x)^2}$, i.e.,
\begin{equation*}
A_{ij} \sim \mathcal{N}\Big(0,\frac{P_T}{n^2 d^2 (d \sigma_x + \mu_x)^2}\Big),
\end{equation*}
where $\sigma_x^2$ and $\mu_x$ are the statistical variance and mean values of the source signals $\mathbf{x} \in \mathbb{R}^n$, respectively, and $d > 0$ is a real number derived from the Chebyshev's inequality.
\label{prop:main}
\end{proposition}

Due to the limited space, please refer to the proof of Proposition 1 and the detailed parameters in Appendix A of our extended version of this work \cite{hieu2024reconstructing}. In particular, Proposition 1 proposes the new construction of measurement matrix $\mathbf{A}$ that satisfies the S-REC property and power constraint at the same time. Our proof in \cite{hieu2024reconstructing} shows that the upper and lower bounds of elements $A_{ij}$ of $\mathbf{A}$ can be derived by applying a series of inequalities, i.e., Bernstein inequality, generalized triangle inequality, and Chebyshev's inequality.
Once the matrix $\mathbf{A}$ is constructed based on Proposition 1, the power constraint (\ref{eq:problem-main}b) can be reduced. Thus, the optimization problem (\ref{prop:main}) is equivalent to solving the Lagrangian of $\mathcal{P}_2$, for some parameter $\lambda \geq 0$, i.e.,
\begin{equation}
\label{eq:gz-lagrange}
\mathcal{P}_3: \quad \min_{\mathbf{z}} \|\mathbf{A}G(\mathbf{z}) - \mathbf{\hat{y}}\|_2^2 + \lambda \|G(\mathbf{z})\|_1.
\end{equation}
As $\mathbf{z}$ is differentiable with respect to the VAE's parameters, one can use the loss function based on (\ref{eq:gz-lagrange}) to train the VAE model. Let $\mathbf{z}^*$ denote the output solution of $\mathcal{P}_3$, the reconstruction error can be bounded with probability $1 - e^{-\Omega(m)}$ by  $\|G(\mathbf{z}) - \mathbf{x}^*\|)_2 \leq 6 \min_{\mathbf{z} \in \mathbb{R}^k} \|G(\mathbf{z}) - \mathbf{x}\|_2 + 3 \|\bm{\eta}\|_2 + 2 \epsilon$,
where $\epsilon$ is an error term caused by the gradient descent optimizer \cite{bora2017compressed}. 

%\begin{figure}
%\centering
%\includegraphics[width=0.75\linewidth]{figures/network-architecture.pdf}
%\caption{Network architecture of the VAE model employed at the receiver.}
%\label{fig:network-architecture}
%\end{figure}

We depict our network architect for the VAE model as follows. The input of the VAE's model is the received signal $\mathbf{\hat{y}}$ at the transmitter, followed by a hidden layer having $64$ neurons. The latent space $\mathbf{z}$ is another hidden layer with the size of $10$-neuron.  The VAE's decoder is a hidden layer with the size $64$-neuron and the output layer is $204$, which is equivalent to the dimension of the original signals. ReLU is selected as the activation function for the hidden layers. Tanh is applied at the output layer as our dataset has normalized IMU signals within the range (-1, 1) (as we will describe later). From hereafter, we refer to our proposed framework as ``CS-VAE" (Compressive Sensing-based Variational Auto-Encoder) for further evaluation.

\vspace{-0.1cm}
\section{Performance Evaluation}
\subsubsection{Dataset and Simulation Settings}
In this section, we validate our proposed CS-VAE framework on the real-world dataset of IMU signals, named DIP-IMU dataset \cite{huang2018deep}. The DIP-IMU dataset contains 330,178 IMU data frames from 17 IMU sensors placed on the human participants. The sampling rate of the IMU sensors is 60 frames per second. The participants in the experiments are asked to perform various actions, categorized into different action classes, e.g., locomotion, freestyle, upper body, and lower body movements. Detailed of the action classes and corresponding number of data frames can be found in \cite{huang2018deep}. After removing abnormal data points (i.e., data points with NaN values), we have the training set and test set containing 220,760 and 56,990 data frames, respectively. We normalize the training set and test set within the range (-1, 1). Each data frame in the dataset, denoted by $\mathbf{x}_t$ has 204 features of acceleration and orientation of the 17 IMU sensors, i.e., $\mathbf{x}_t \in \mathbb{R}^{204}$. We train the VAE model for 50 epochs with batch size 60. We use Adam optimizer with a learning rate of $10^{-4}$ to train the VAE model. The $l_1$ penalty in (\ref{eq:gz-lagrange}) is $10^{-5}$.

We evaluate the proposed CS-VAE framework and compare it with two baselines that are (i) Lasso (Lease absolute shrinkage and selection operator) \cite{foucart2013invitation} and (ii) DIP (Deep Inertial Poser) \cite{huang2018deep}. In particular, Lasso is a classical regression technique for finding the sparse solution in compressive sensing problems. In our evaluation, Lasso finds the solution for equation (\ref{eq:lasso-lagrangian}) with a similar $l_1$ penalty as the CS-VAE's, i.e., $\lambda = 10^{-5}$. With the DIP baseline, we use the same approach in \cite{huang2018deep} in which we manually select the IMU sensors for down-sampling the IMU signals before transmitting them to the receiver. To deal with the power constraint (\ref{eq:problem-main}b), we apply a $l_2$ normalization in which the real-valued vector $\mathbf{y}$ is divided to its $l_2$ vector norm, i.e., $\frac{1}{\|\mathbf{y}\|_2} \mathbf{y}$. Although the $l_2$ normalization is widely used due to its simplicity, we later show that it yields a non-linear transform of $\mathbf{y}$, which makes the receiver unable to guarantee the high accuracy of the reconstructed signal. Our proposed novel measurement matrix design does not suffer from this problem, as examined in the following.

\subsubsection{Simulation Results}

\begin{figure}
\centering
\includegraphics[width=0.55\linewidth]{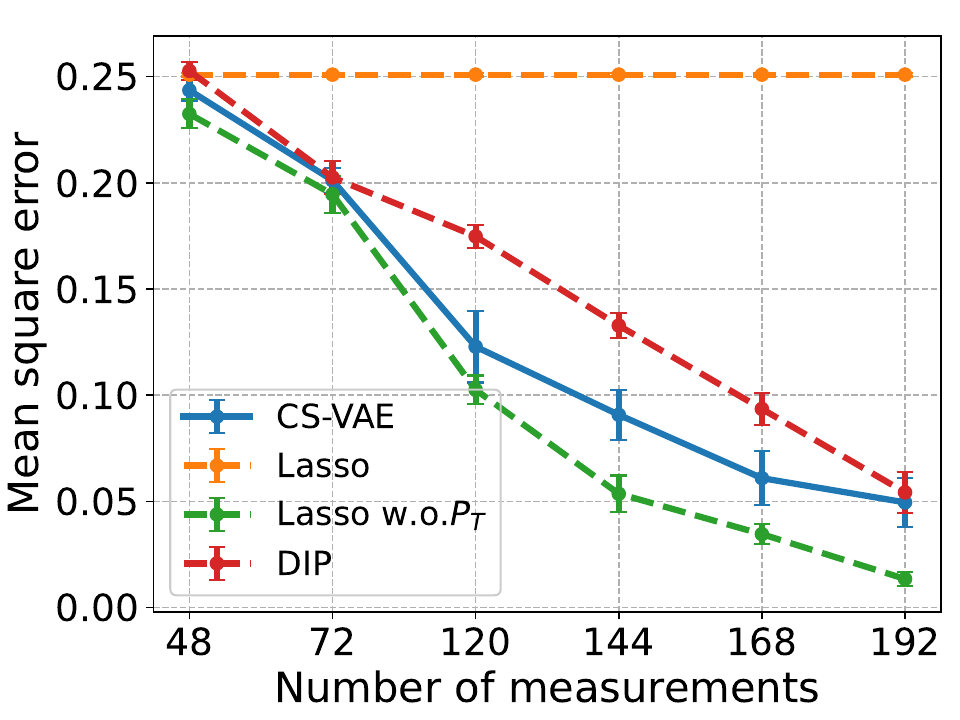}
\caption{Reconstruction accuracy vs. number of measurements.}
\label{fig:mn-result}
\end{figure}

We first evaluate the impacts of the number of measurements $m$ on the system performance. We use the mean square error value (MSE) of the reconstructed signal as our main evaluation metric. In Fig.~\ref{fig:mn-result}, we vary the number of measurements $m$ from 48 to 192. The ``Lasso" baseline in Fig.~\ref{fig:mn-result} is the Lasso baseline with the use of measurement matrix $\mathbf{A}$ in Proposition \ref{prop:main} to ensure the power constraint. The ``Lasso w.o.$P_T$" baseline denotes the Lasso approach with measurement matrix $\mathbf{B}$ with elements $B_{ij} \sim \mathcal{N}(0, 1/m)$, similar to the setting in \cite{bora2017compressed} and \cite{dhar2018modeling}. The ``Lasso w.o.$P_T$" baseline is the Lasso optimizer without considering the power constraint. In our evaluation, this baseline serves as the upper bound as it is not subjected to the power constraint in (\ref{eq:problem-main}b), resulting in possible higher signal-to-noise ratios. It can be observed from Fig.~\ref{fig:mn-result} that the increase of $m$ results in decreasing the MSE values. The proposed CS-VAE approach obtains lower MSE values than the DIP and Lasso baseline in all scenarios. With the Lasso baseline, its MSE values remain unchanged with the increase of $m$, meaning that it fails to reconstruct the original signal $\mathbf{x}^*$ from the observation $\mathbf{\hat{y}}$. The Lasso w.o.$P_T$ baseline produces the best performance as the power constraint does not hold. This result reveals that the impact of the power constraint on the system and our proposed CS-VAE framework can achieve significant improvement in reconstruction accuracy. 

\begin{figure}
\centering
\includegraphics[width=0.55\linewidth]{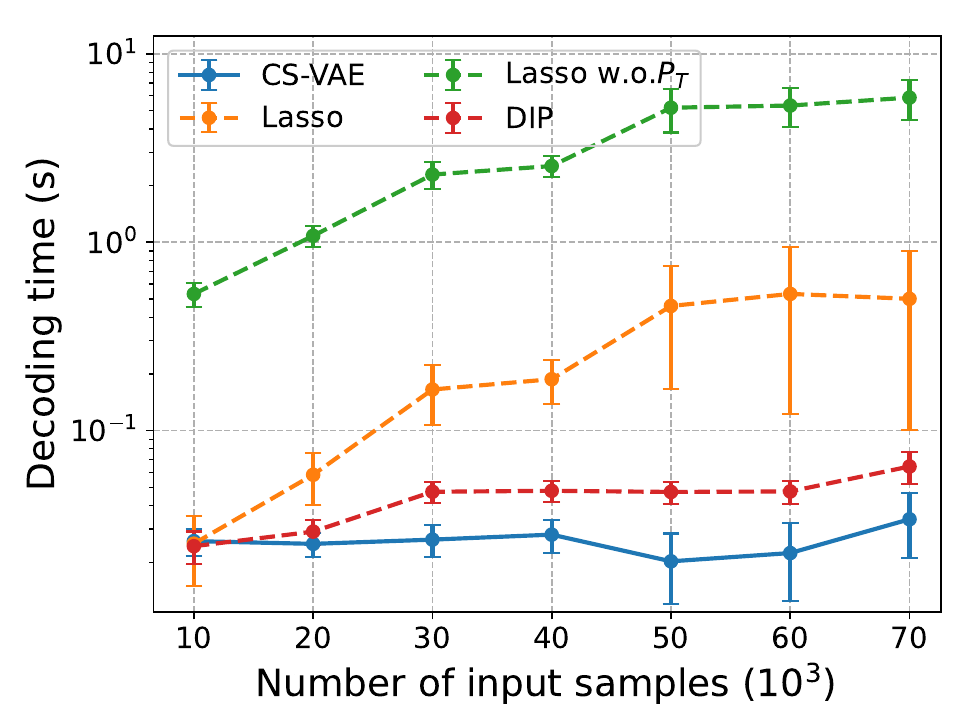}
\caption{Reconstruction latency vs. number of inputs.}
\label{fig:nb-result}
\end{figure}

Next, we evaluate the impact of the number of received samples (or the input samples of the VAE model) at the receiver on the decoding latency of the receiver. The decoding latency is defined as the time from receiving the compressed signals to the time the original signals are entirely reconstructed. This number of input samples in Fig.~\ref{fig:nb-result} is equivalent to $m \times b$, where $m$ is the number of measurements and $b$ is the batch size of the input samples. As observed from Fig.~\ref{fig:nb-result}, the decoding time of the Lasso baselines increases with the number of input samples. Notably, the high accuracy of the Lasso w.o.$P_T$ baseline comes with significant latency as it requires a longer time to find the solution for the problem defined in (\ref{eq:lasso-lagrangian}). Our proposed CS-VAE approach is more suitable for large batches of input samples as the pre-trained VAE model can reconstruct the signals through a single forward pass across the hidden layers, resulting in a magnitude faster in decoding latency with comparable accuracy to the optimal solution obtained by the Lasso w.o.$P_T$ baseline.

\begin{figure}
\centering
\includegraphics[width=0.7\linewidth]{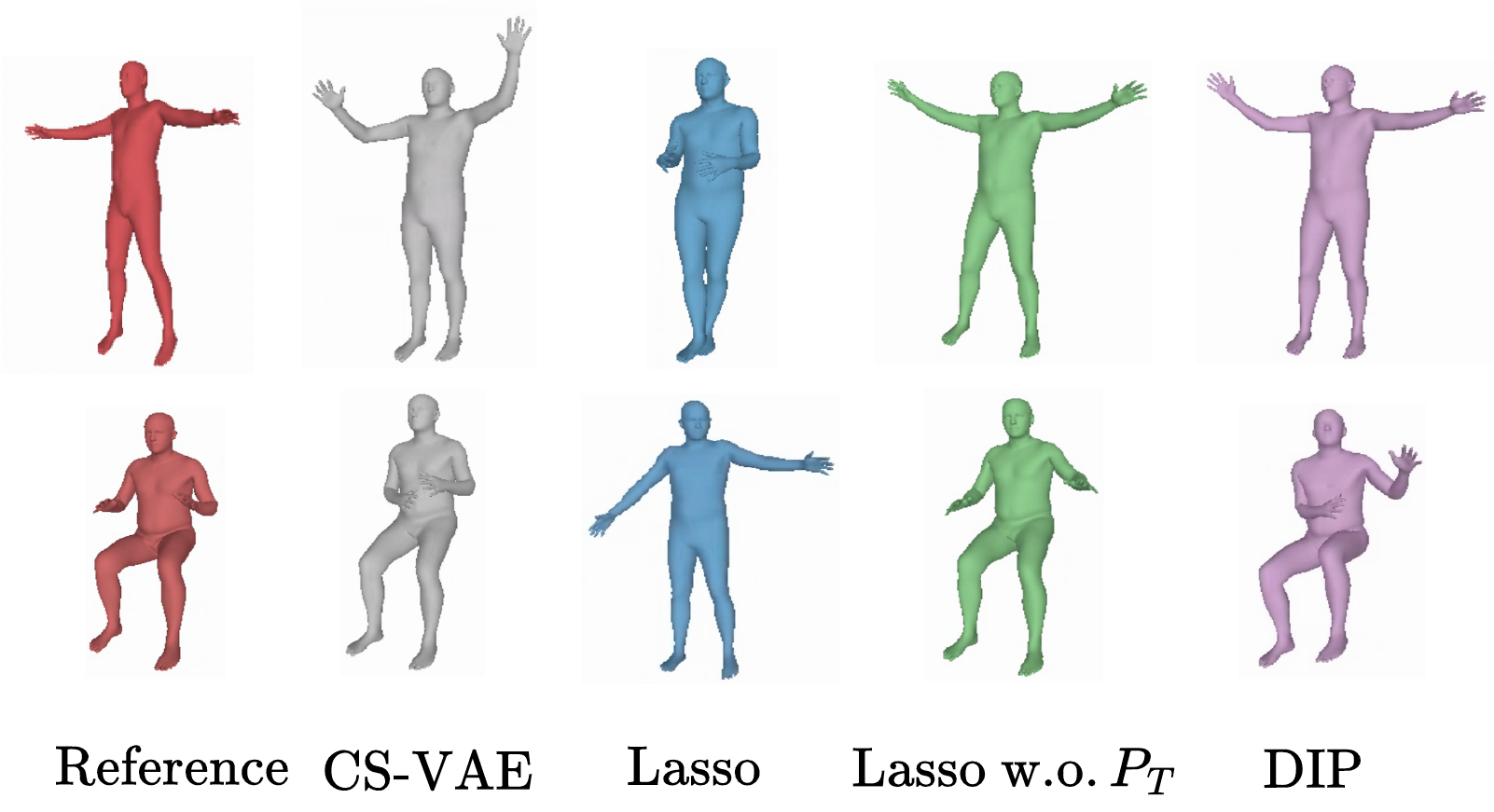}
\caption{3D reconstruction poses based on the reconstructed IMU signals.}
\label{fig:pose-reconstruction}
\end{figure}

Finally, we use the reconstructed signals at the receiver to produce the 3D human movements, as illustrated in Fig.~\ref{fig:pose-reconstruction}. To obtain the 3D human models in Fig.~\ref{fig:pose-reconstruction}, we use another pre-trained VAE model to map the 204-dimensional signal $\mathbf{\hat{x}}$ into a 72-dimensional vector $\mathbf{p}$ representing the pose of the human avatar. In other words, the pre-trained VAE model serves as a mapping function $F(\mathbf{\hat{x}}): \mathbb{R}^{204} \rightarrow \mathbb{R}^{72}$. The pose parameters $\mathbf{p} \in \mathbb{R}^{72}$ are input of a kinematic 3D human model, named SMPL \cite{loper2023smpl}. The details of the mapping function $F(\mathbf{\hat{x}})$ with the VAE model can be found in our extended version of this work \cite{hieu2024reconstructing}. As observed from Fig.~\ref{fig:pose-reconstruction}, our proposed CS-VAE can generate precise 3D poses in different samples from the test set with $m=168$ measurements (i.e., $82\%$ of the original signal). Similar to the above analyses, the upper bound baseline Lasso w.o.$P_T$ generates the most accurate 3D poses, followed by our proposed CS-VAE framework, while the Lasso baseline fails to reconstruct the 3D human poses.

\vspace{-0.1cm}
\section{Conclusion}
In this paper, we have developed a novel generative model-based compressive sensing framework for the reconstruction of 3D human poses from sparse IMU measurements. The proposed measurement matrix design enabled the generative model, i.e., a variational auto-encoder, to recover the original IMU signals from compressed and noisy received signals. The principle idea of this work is designing the measurement matrix that ensures the power constraint of the transmitter over an additive white Gaussian noise channel and also guarantees the S-REC property in compressive sensing. Simulation results on the real-world dataset DIP-IMU have shown the best performance of the proposed framework, compared with other baselines. From the recovered signals at the receiver, we have shown that it is possible to recover the full 3D human body poses mimicking the movements of the user. This paves the way for a wide variety of Metaverse applications in which estimating the 3D body movements of the user plays an important role. Potential approaches extending from this work could be extending our matrix design to more complicated channels, such as fading channels, or multiple-access scenarios.

\vspace{-0.1cm}
\bibliographystyle{IEEEtran}
\bibliography{references}

\begin{thebibliography}{10}
\providecommand{\url}[1]{#1}
\csname url@samestyle\endcsname
\providecommand{\newblock}{\relax}
\providecommand{\bibinfo}[2]{#2}
\providecommand{\BIBentrySTDinterwordspacing}{\spaceskip=0pt\relax}
\providecommand{\BIBentryALTinterwordstretchfactor}{4}
\providecommand{\BIBentryALTinterwordspacing}{\spaceskip=\fontdimen2\font plus
\BIBentryALTinterwordstretchfactor\fontdimen3\font minus
  \fontdimen4\font\relax}
\providecommand{\BIBforeignlanguage}[2]{{%
\expandafter\ifx\csname l@#1\endcsname\relax
\typeout{** WARNING: IEEEtran.bst: No hyphenation pattern has been}%
\typeout{** loaded for the language `#1'. Using the pattern for}%
\typeout{** the default language instead.}%
\else
\language=\csname l@#1\endcsname
\fi
#2}}
\providecommand{\BIBdecl}{\relax}
\BIBdecl

\bibitem{siriwardhana2021survey}
Y.~Siriwardhana \emph{et~al.}, ``A survey on mobile augmented reality with 5g
  mobile edge computing: Architectures, applications, and technical aspects,''
  \emph{IEEE Commun. Surv. \& Tut.}, vol.~23, no.~2, pp. 1160--1192, Feb. 2021.

\bibitem{5gamericas2024}
\BIBentryALTinterwordspacing
5G-Americas, ``3gpp technology trends,'' Jan. 2024, white paper. [Online].
  Available: \url{https://www.5gamericas.org/3gp-technology-trends/}
\BIBentrySTDinterwordspacing

\bibitem{guo2023federated}
Y.~Guo, Z.~Qin, X.~Tao, and G.~Y. Li, ``Federated multi-view synthesizing for
  metaverse,'' \emph{IEEE J. on Sel. Areas in Commun.}, vol.~42, no.~4, Apr.
  2024.

\bibitem{jiang2023qoe}
Y.~Jiang, J.~Kang, X.~Ge, D.~Niyato, and Z.~Xiong, ``Qoe analysis and resource
  allocation for wireless metaverse services,'' \emph{IEEE Trans. on Commun.},
  2023.

\bibitem{fernandes2016combating}
A.~S. Fernandes and S.~K. Feiner, ``Combating vr sickness through subtle
  dynamic field-of-view modification,'' in \emph{IEEE Symp. on 3D User
  Interfaces}, Mar. 2016, pp. 201--210.

\bibitem{xsens2023}
\BIBentryALTinterwordspacing
Movella, ``Superior xsens motion capture technology, optimized for human
  movement,'' accessed: Apr. 4. [Online]. Available:
  \url{https://www.movella.com/products/wearables/xsens-mtw-awinda}
\BIBentrySTDinterwordspacing

\bibitem{von2017sparse}
T.~Von~Marcard, B.~Rosenhahn, M.~J. Black, and G.~Pons-Moll, ``Sparse inertial
  poser: Automatic 3d human pose estimation from sparse imus,'' in
  \emph{Comput. Graph. Forum}, vol.~36, no.~2, May 2017, pp. 349--360.

\bibitem{huang2018deep}
Y.~Huang \emph{et~al.}, ``Deep inertial poser: Learning to reconstruct human
  pose from sparse inertial measurements in real time,'' \emph{ACM Trans. on
  Graph.}, vol.~37, no.~6, pp. 1--15, Dec. 2018.

\bibitem{yi2022physical}
X.~Yi \emph{et~al.}, ``Physical inertial poser (pip): Physics-aware real-time
  human motion tracking from sparse inertial sensors,'' in \emph{Proc. of the
  IEEE/CVF Conf. on Comput. Vision and Pattern Recognit.}, Jun. 2022, pp.
  13\,167--13\,178.

\bibitem{loper2023smpl}
M.~Loper \emph{et~al.}, ``Smpl: A skinned multi-person linear model,'' in
  \emph{Seminal Graphics Papers: Pushing the Boundaries, Volume 2}, 2023, pp.
  851--866.

\bibitem{bora2017compressed}
A.~Bora, A.~Jalal, E.~Price, and A.~G. Dimakis, ``Compressed sensing using
  generative models,'' in \emph{Int. Conf. on Mach. Learn.}, Jul. 2017, pp.
  537--546.

\bibitem{foucart2013invitation}
S.~Foucart and H.~Rauhut, \emph{A Mathematical Introduction to Compressive
  Sensing}.\hskip 1em plus 0.5em minus 0.4em\relax Springer, 2013.

\bibitem{raskutti2010restricted}
G.~Raskutti, M.~J. Wainwright, and B.~Yu, ``Restricted eigenvalue properties
  for correlated gaussian designs,'' \emph{The Journal of Machine Learning
  Research}, vol.~11, pp. 2241--2259, 2010.

\bibitem{dhar2018modeling}
M.~Dhar, A.~Grover, and S.~Ermon, ``Modeling sparse deviations for compressed
  sensing using generative models,'' in \emph{Int. Conf. on Mach. Learn.}, Jul.
  2018, pp. 1214--1223.

\bibitem{kingma2013auto}
D.~P. Kingma and M.~Welling, ``Auto-encoding variational bayes,'' in \emph{Int.
  Conf. on Learn. Representations}, Apr. 2014, pp. 1--14.

\bibitem{hieu2024reconstructing}
N.~Q. Hieu, D.~T. Hoang, D.~N. Nguyen, and M.~A. Alsheikh, ``Reconstructing
  human pose from inertial measurements: A generative model-based compressive
  sensing approach,'' \emph{IEEE Journal on Selected Areas in Communications},
  Jun. 2024.

\end{thebibliography}
\end{document}